\documentclass[article]{revtex4}

\usepackage{graphicx}
\usepackage{dcolumn}
\usepackage{bm}
\usepackage{color}

\usepackage{eso-pic}
\usepackage{amsmath }
\usepackage{latexsym}
\usepackage[french]{babel}

\begin{document}
\large

\date{\today}
\title{Two-dimensional superconductivity at a Mott-Insulator/Band-Insulator interface: $\mbox{LaTiO}_3$/$\mbox{SrTiO}_3$ }

\author{J. Biscaras$^1$, N. Bergeal$^1$, A. Kushwaha$^2$, T. Wolf$^1$,  A. Rastogi$^2$, R.C. Budhani$^{2,3}$  and J. Lesueur$^1$}

\affiliation{$^1$LPEM- UMR8213/CNRS - ESPCI  ParisTech, 10 rue Vauquelin - 75005 Paris}

\affiliation{$^2$Condensed Matter - Low Dimensional Systems Laboratory, Department of Physics, Indian Institute of Technology Kanpur, Kanpur 208016, India}
\affiliation{$^3$National Physical Laboratory, New Delhi - 110012, India.}

\maketitle

Transition metal oxides display a great variety of quantum electronic
behaviours where correlations often play an important role. The achievement
of high quality epitaxial interfaces involving such materials gives a unique opportunity to engineer artificial structures where new electronic orders take place. One of the most striking result in this area is the recent observation of a 
two-dimensional electron gas at the interface between  a strongly correlated Mott insulator $\mbox{LaTiO}_3$ and a band insulator $\mbox{SrTiO}_3$  \cite{ohtomo02,ohtomo04}. The mechanism responsible for such a behaviour is still under debate. In particular, the influence of the nature of the insulator has to be clarified. Here we show that despite the expected electronic correlations, $\mbox{LaTiO}_3$/$\mbox{SrTiO}_3$ heterostructures undergo a superconducting transition at a critical temperature $T^{\mathrm{onset}}_c\sim$ 300 mK. We have found that the superconducting electron gas is confined over a typical thickness of 12 nm. We discuss the electronic properties of this system and review the possible scenarios. \\

Perovskites based structures including transition metal oxides have
attracted much attention in the last decades, with the discovery of
high-T$_c$ superconductivity and colossal magnetoresistance \cite{dagotto}.
More generally, these compounds exhibit  various electronic orders,
going from the canonical Anti-Ferromagnetic (AF) Mott insulator when
the on-site repulsion is maximum because of strong electronic interactions,
to Fermi Liquid like metals when carrier doping is such that screening
prevents the system from localisation. Depending on the cations and
the doping level involved, charge, spin and orbital orders
can appear in the ground state together with metallic and even superconducting
phases. Transitions between these states can be tuned by temperature,
magnetic or electric fields \cite{imada}. All these 
compounds can be seen as stacks of oxide layers where the charge neutrality
is conserved in the unit cell, but not necessarily in each layer.
Therefore, the translation symmetry is locally broken at the interface  and charge imbalance can
develop. Like in band gap engineering with semiconductors, it is possible to create artificial interface materials by growing thin layers
of a transition metal oxide on top of another one. Recently, the observation of two-dimensional superconductivity  and magnetic correlations \cite{brinkman} at the interface between the two band insulators $\mbox{LaAlO}_3$ and $\mbox{SrTiO}_3$ \cite{reyren} has drawn a lot of attention. An other particularly interesting candidate is the homo-metallic structure  $\mbox{LaTiO}_3$/$\mbox{SrTiO}_3$ that uses $\mbox{TiO}_2$  plans as a building block \cite{ohtomo02}.
Titanium is in the $3d^{0}$ state in the $\mbox{SrTiO}_3$  layer which is a band insulator
of 3.2 eV bandgap, while it is $3d^{1}$ in the $\mbox{LaTiO}_3$  one which is
therefore an AF Mott insulator due to strong correlations
\cite{tokura92}. Providing the interface layer is $\mbox{TiO}_2$, an extra electron
 is left in the structure every two unit cells \cite{okamoto04n,okamoto04p}. As shown by photoemission \cite{takizawa}
and optical studies \cite{seo}, a two-dimensionnal electron
gas (2-DEG) develops  and extends a few unit cells beyond the interface.\\

   \begin{figure}[htpb]
\centering
\includegraphics[width=10cm]{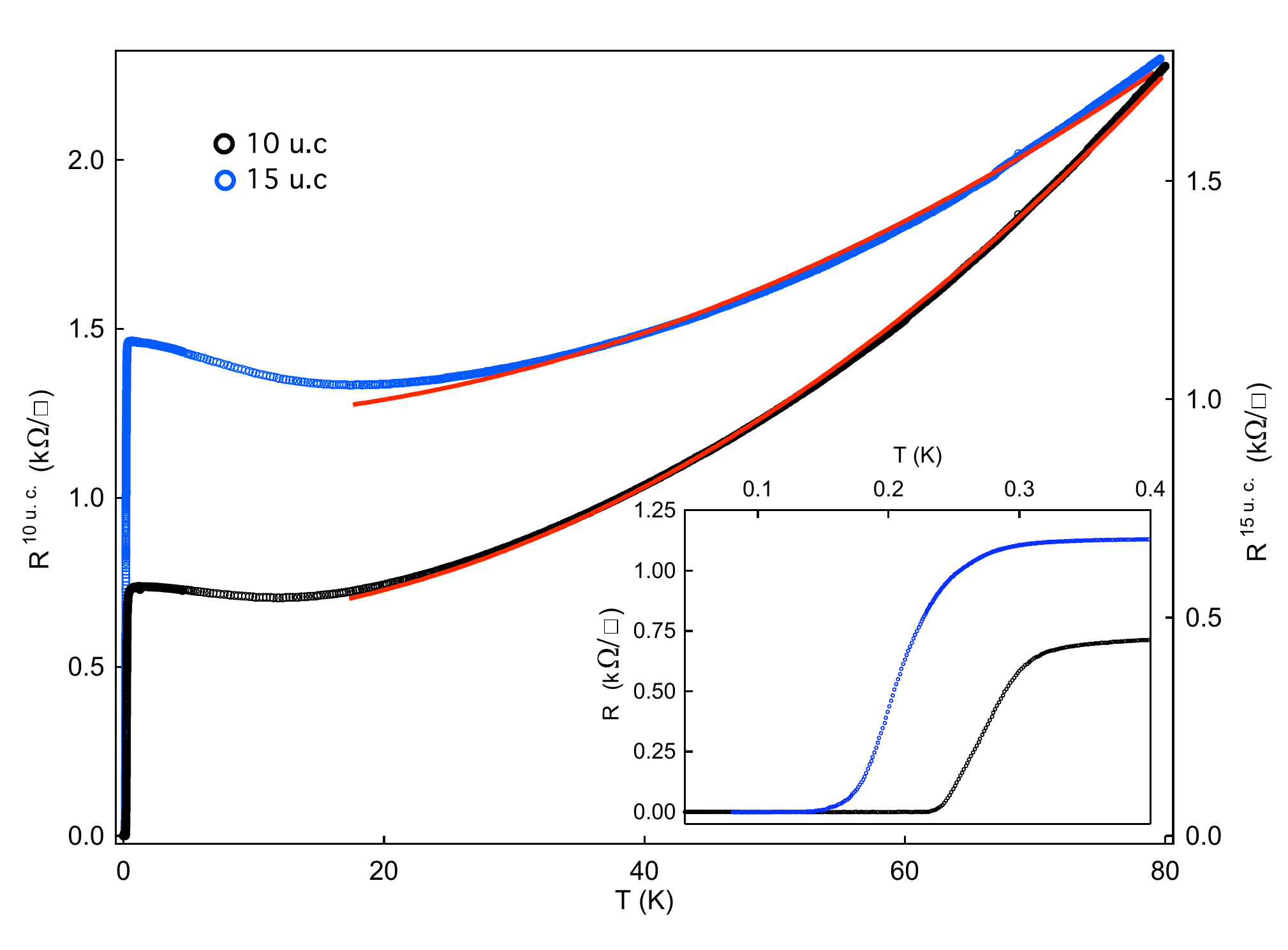}
\caption{Sheet resistance a of the 10 u.c. (left axis) and 15 u.c  (right axis) $\mbox{LaTiO}_3$/$\mbox{SrTiO}_3$ samples in an intermediate range of temperature.The red lines correspond to quadratic fits of the form $R(T)=AT^2+R_0$. Inset ) Sheet resistance as a function of temperature showing the superconducting transitions at $T_c^{\mathrm{onset}}\approx$310mK for the  10 u.c. and  $T_c^{\mathrm{onset}}\approx$260mK for the 15 u.c., where $T_c^{\mathrm{onset}}$ is defined by a 10$\%$ drop of the resistance.\\}
\end{figure}

 Several theoretical approaches pointed out that  an electronic reconstruction leads to an increase of the electronic density at the $\mbox{LaTiO}_3$/$\mbox{SrTiO}_3$ interface \cite{okamoto04n,ishida,okamoto04p,popovic}. Okamoto and Millis proposed a phase diagram where different orbital
and magnetic states occur as a function of the thickness of the $\mbox{LaTiO}_3$ 
layer and the strength of the Mott-Hubbard parameter $U/t$ ($U$ is the
Coulomb on-site repulsion energy, and $t$ the hopping term between neighbour 
Ti sites)\cite{okamoto04n}. Fully polarized ferromagnetic metallic sub-bands are expected
to form for thickness below five unit cells and $U/t\sim8-10$. On the
other hand Kancharla and Dagotto \cite{kancharla} taking
into account both local and long range Coulomb interactions, showed
that strong AF fluctuations reminiscent of the magnetic order of the
bulk compound persist in the metallic phase. As suggested by Larson \cite{larson} and Okamoto \cite{okamoto06},  lattice relaxation
at the interface strongly modifies the band configuration,
and may enhance the electronic correlations in the 2-DEG \cite{ishida}. In
this context, it is clear that the $\mbox{LaTiO}_3$/$\mbox{SrTiO}_3$  interface layer appears
to be a unique system to study the physics of a 2-DEG influenced by strong electronic correlations.

   \begin{figure}[htpb]
\centering
\includegraphics[width=10cm]{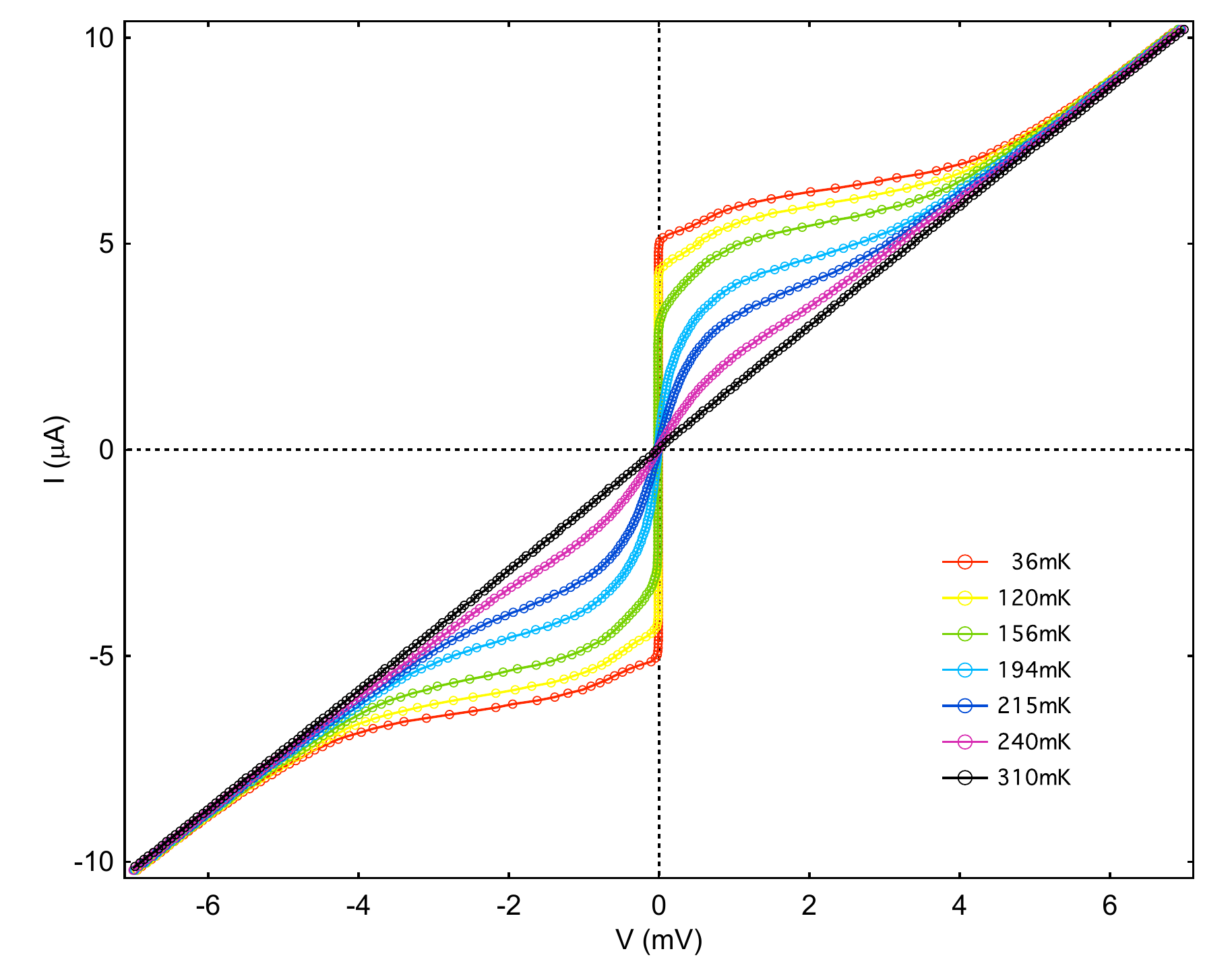}
\caption{Current-voltage characteristics of the 10 u.c.  sample at different temperatures. The critical current at low temperature is 5  $\mu$A corresponding to a critical current per unit width of 16.7 $\mu$A/cm.}
\end{figure}

We have grown epitaxial layers of $\mbox{LaTiO}_3$ using
excimer laser based Pulsed Laser Deposition (PLD) on  single crystal substrates of $\mbox{SrTiO}_3$
cut along (100) and (110) crystallographic directions. The details of the growth conditions and X-rays characterisations
 are given in Supplementary Information. In this study, we focus mainly on two $\mbox{LaTiO}_3$/(100)$\mbox{SrTiO}_3$ hetero-structures whose thickness  of 40 $\AA$ and 60 $\AA$ correspond to 10  and 15 unit cells (u.c.) respectively.   The sheet resistance measured in a Van-der-Pauw geometry decreases with temperature, indicating a metallic behaviour of the interface (figure 1). At temperatures lower than 20K the two samples exhibit an increase of resistance characteristic of weak localisation in disordered two-dimensional films. The hetero-structures undergo a superconducting transition at $T_c^{\mathrm{onset}}\approx$ 310 mK for the 10 u.c. sample and at $T_c^{\mathrm{onset}}\approx$ 260 mK for the 15 u.c.  sample  (inset figure1). Thinner 5 u.c. (100) films and  20 to 100 u.c. thicker (100) films as well as  (110) oriented films are not metallic at low temperature.\\
 
   \begin{figure}[h!]
\centering
\includegraphics[width=10cm]{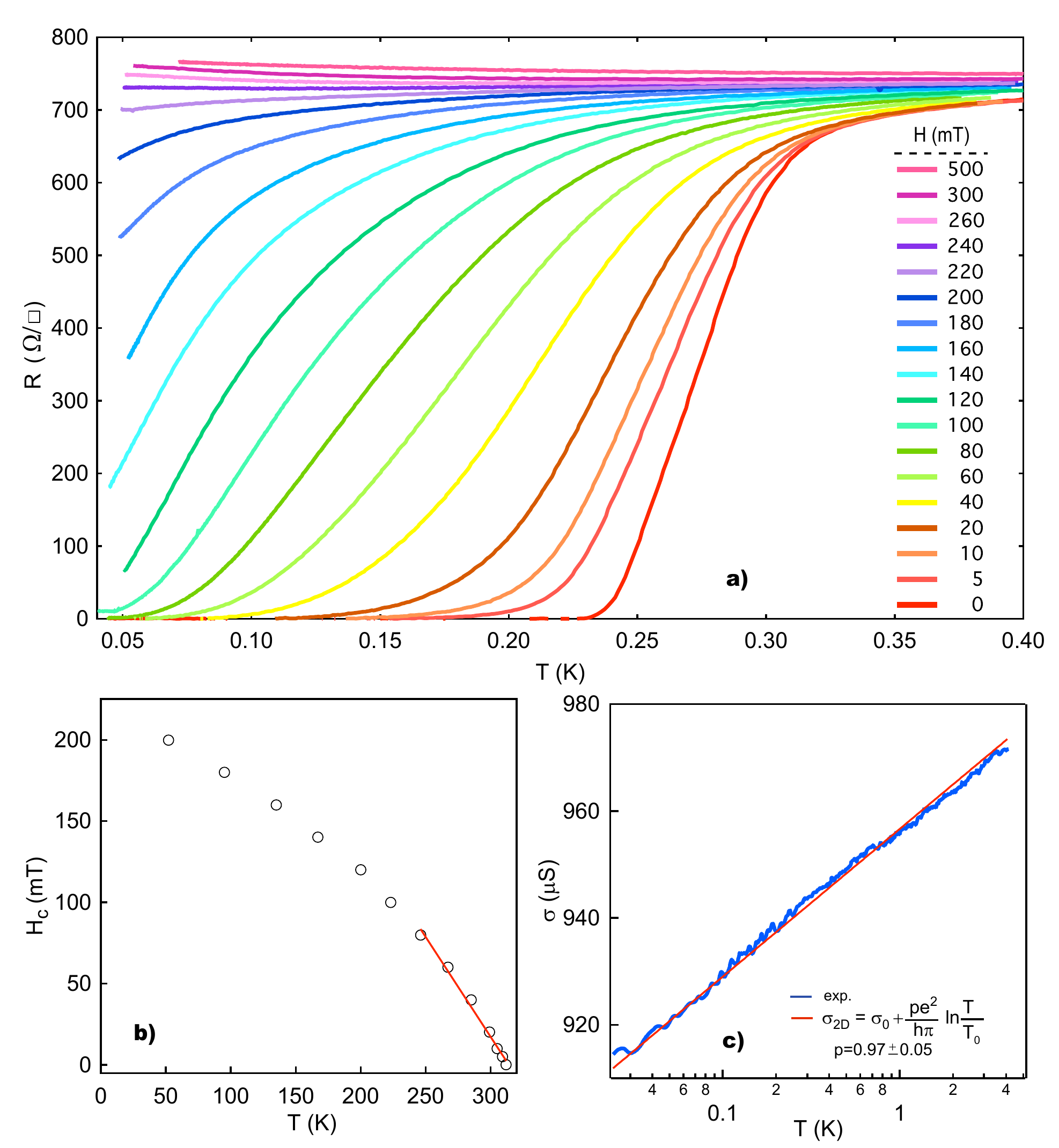}
\caption{a) Sheet resistance of the 10 u.c. sample as a function of temperature for different values of the perpendicular magnetic field. b) Temperature dependence of the perpendicular critical field, defined  as the magnetic field that suppresses 90$\%$ of the resistance drop. The red line indicates the linear dependence of the critical field with temperature close to $T_c$. c) Conductivity of the 15 u.c. sample as a function of temperature for a perpendicular magnetic field corresponding to the critical field. The red line corresponds to the expression $\sigma_{2D}(T)=\sigma_0+\frac{pe^2}{\pi h} \ln{\frac{T}{T_0}}$  with $p=0.97$ indicating that phase coherence is limited by electron-electron scattering ($p$=1)\cite{lee}.}
\end{figure}

  In figure 2, we show the current-voltage characteristics of the 10 u.c. sample measured at different temperatures. At low temperature, the I(V) curves displays a clear critical current $I_c$ of 5  $\mu$A corresponding to a critical current per unit width of 16.7 $\mu$A/cm. For current much higher than $I_c$, the I(V) curves merge together on a linear Ohmic law with a resistance corresponding to the normal resistance. In the case of the 15 u.c. sample, the critical current per unit width is found to be 14 $\mu$A/cm. Figure 3a shows the sheet resistance of the 10 u.c sample as a function of temperature measured for different values of a magnetic field applied perpendicularly to the sample. The magnetic field induces a transition from a superconducting state to a non-superconducting one. The dependence of the critical field as a function of temperature is linear close to  $T_c$, which is consistent with the form $H(T)=\frac{\Phi_0}{2\pi\xi_\parallel^2(T)}$ taking into account a Landau-Ginsburg in-plane coherence length $\xi_\parallel\propto(T_c-T)^{-\frac{1}{2}}$ (figure 3b). The critical field extrapolated at $T$=0 is  $H_{\perp}^c\approx$ 220 mT for the 10 u.c. sample and $H_{\perp}^c\approx$ 210 mT for the 15 u.c. sample (see supplementary figure 3). At T=0, we found $\xi^{10 u.c.}_\parallel(T=0)\approx$ 38 nm  and $\xi^{15u.c.}_\parallel(T=0)\approx$ 42 nm.  Measurements performed in a parallel magnetic field geometry give $H_{\parallel}^c=\frac{\sqrt{3}\Phi_0}{\pi d \xi_{\parallel}(T=0)}\approx$ 2.15 T for the 10 u.c sample and $H_{\parallel}^c$=2.2 T for the 15 u.c sample.  We thus extract the thickness of the two-dimensional superconducting electron gas $d^{15u.c.}\approx$ 12 nm  and  $d^{10u.c.}\approx$ 13.5 nm. Note that this is an upper bound given the  precision of the sample alignment in the parallel magnetic field. These values are closed to the ones reported in $\mbox{LaAlO}_3$/$\mbox{SrTiO}_3$ hetero-structures\cite{basletic,copie}. In disordered electronic system, weak localisation produces a decrease of  conductivity that can be experimentally revealed by varying the temperature.  In the particular case of a two-dimensional system, the conductivity takes the remarkable logarithmic dependence with temperature  $\sigma_{2D}(T)=\sigma_0+\frac{pe^2}{\pi h} \ln{\frac{T}{T_0}}$ where $p$ depends on the process that limits the phase coherence. $p=3$ for electron-phonon scattering and $p=1$ for electron-electron scattering in the dirty limit\cite{lee}. Such logarithmic temperature dependence is observed on our samples (see figure 3c), thus confirming the two-dimensional nature of the electron gas. The fit gives $p=0.97\pm0.05$ showing that the phase coherence is limited mainly by electron-electron scattering.\\
  
 It is known that $\mbox{LaTiO}_3$ itself can be oxygen \cite{taguchi, gariglio,wang}
or Sr doped \cite{tokura}, and thus becomes metallic. The
key question is therefore : does superconductivity take place within
a doped Mott insulator layer, namely oxygen or Sr doped $\mbox{LaTiO}_3$, or
within a 2-DEG formed at the $\mbox{LaTiO}_3$/$\mbox{SrTiO}_3$   interface, which extends 
mostly within the band insulator $\mbox{SrTiO}_3$ \cite{okamoto04n} ?
And conversely, do the electronic correlations, which are known
to be strong in the former case and moderate in the latter one\cite{okamoto06},
play a role in that context ? The recent works on $\mbox{LaTiO}_3$/$\mbox{SrTiO}_3$ superlattices \cite{ohtomo02,seo,shibuya}
clearly indicate that under proper growth conditions, the interface
is abrupt, with no sizable Sr diffusion for deposition temperatures below 1000 $^\circ$C\cite{takizawa}.
Optical spectroscopy  confirms that the carrier
properties in superlattices are different than the ones in $\mbox{La}_{1-x}$$\mbox{Sr}_x$$\mbox{TiO}_3$ compounds \cite{seo}.
Moreover, the low temperature transport properties are different in conducting  doped $\mbox{LaTiO}_3$  and doped $\mbox{SrTiO}_3$. In both
cases, electron-electron collisions dominate the scattering events
according to the Fermi liquid picture.  Figure 1 shows that, in an intermediate regime temperature, the temperature dependence of the resistance is well fitted by a quadratic law $R(T)=AT^2+R_0$ where the coefficient $A$ depends on the Landau parameters, and therefore on the carrier density and the effective mass m{*}\cite{nozieres}. We obtained $A$=0.27 $\Omega/\Box K^{2}$ for the 10 u.c. sample and $B$=0.11 $\Omega/ \Box K^{2}$ for the  15 u.c. sample. In table [1] we summarise the different values of $A$ found in the literature for  doped $\mbox{LaTiO}_3$ and doped $\mbox{SrTiO}_3$ thin films and crystals and compare them to  the values extracted from our experiment. The largest value of $A$ reported in doped $\mbox{LaTiO}_3$ are on the order of $1\times10^{-9}\Omega cm/K^{2}$  before the system becomes insulating at low temperature
($\partial R/\partial T<0$ for $T<$100 K), whereas it is  two orders of magnitude higher for doped $\mbox{SrTiO}_3$. From the comparison, we see that the $\mbox{LaTiO}_3$/$\mbox{SrTiO}_3$ interface layer behave more like doped $\mbox{SrTiO}_3$  than doped $\mbox{LaTiO}_3$. This indicates that the 2-DEG extends mostly in the $\mbox{SrTiO}_3$  substrate and makes it conducting \cite{okamoto04n}. \\

   \begin{figure}[h!]
\centering
\includegraphics[width=10cm]{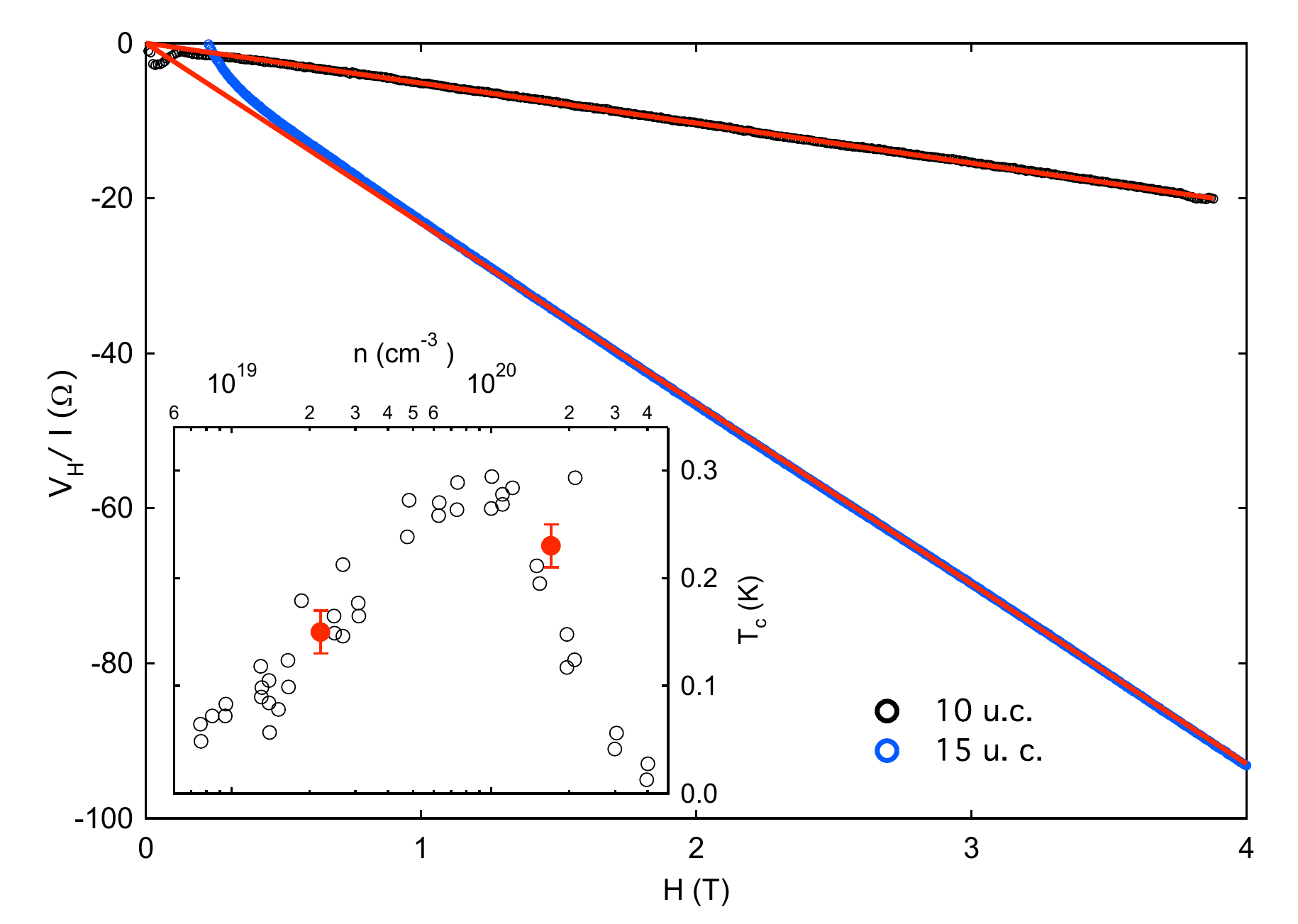}
\caption{a) Hall voltage $V_H$ divided by the current $I$ as a function of magnetic field for the 10 u. c. and 15 u.c.  $\mbox{LaTiO}_3$/$\mbox{SrTiO}_3$ samples, measured at 100 mK. Red solid lines correspond to linear fits. Inset) $T_c$ as a function of doping for SrTiO$_3$ single crystals taken from \cite{koonce} (black open circles). The two red dots corresponds to $\mbox{LaTiO}_3$/$\mbox{SrTiO}_3$ samples taking the thickness $d^{10u.c.}\approx$ 12 nm, $d^{15u.c.}\approx$ 13.5 nm. Here, we have defined  $T_c$ as the temperature at which the resistance reaches zero since this definition is more appropriate for comparison with the magnetic definition of the $T_c$ used in reference \cite{koonce}.}
\end{figure}

 To investigate the density and mobility of charge carriers, we have performed Hall measurements at low temperature (figure 4). The experiment confirms that the sign
 of the hall coefficient $R_H=\frac{V_H}{IB}$ is negative for both samples, indicating that electron-like charge carriers dominate the transport. The  sheet carrier density $n_S=\frac{1}{eR_H}$  was found to be $2 \times10^{14} cm^2$ for the 10 u.c. sample and $=2.7 \times10^{13} cm^2$ for the 15 u.c. one.  Taking the sheet resistance measured previously, we obtained a Hall mobility $\mu=\frac{1}{eR_sn_s}$ of $52$ cm$^2$V$^{-1}$$s^{-1}$ for the 10 u.c. sample and 
 $210$ cm$^2$V$^{-1}$s$^{-1}$ for the 15 u.c. one. 
 In an ideal picture, the interface between $\mbox{SrTiO}_3$  and $\mbox{LaTiO}_3$  can be
seen as a Ti ions network, in the $4^{+}$ state in $\mbox{SrTiO}_3$ 
and in the $3^{+}$ one in $\mbox{LaTiO}_3$. Therefore, one electron is left
every two cells on average at the interface, which corresponds to
an areal density of around $3\times10^{14}cm^{-2}$ \cite{okamoto04n,okamoto04p}. This is approximatively
the electron density measured through Hall effect in the 10 u.c. sample ($2\times10^{14}cm^{-2}$), consistent with the value observed in $\mbox{LaTiO}_3$/$\mbox{SrTiO}_3$  superlattices \cite{ohtomo02} by measuring the number
of $Ti^{3^{+}}$ in the vicinity of the interface. Optical studies
confirm that free carriers with densities of around $3\times10^{14}/cm^{2}$
do exist in similar superlattices, with a typical mobility of $35\, cm^{2}/Vs$
and an effective mass $m^{*}\simeq2m_{e}$  \cite{seo}. The mobility  that we measured  on the 10 u.c. sample ($52\, cm^{2}/Vs$)  is
close to this value, which supports an effective mass close to $2m_{e}$. The 15 u.c. has a lower sheet density of $2.7\times10^{13}/cm^{2}$ and a  $T_c^{\mathrm{onset}}$ of only 260 mK. These observations are consistent with an electronic reconstruction of the $\mbox{LaTiO}_3$/$\mbox{SrTiO}_3$  
interface, leading to the formation of a few unit cells thick 2-DEG in the $\mbox{SrTiO}_3$  layer \cite{okamoto04n,okamoto06,larson}. As shown in figure 4 inset, our data are consistent with the dependence of $T_c$ with the carrier density reported in the literature of doped $\mbox{SrTiO}_3$ \cite{koonce}.
\\

\begin{table}

\begin{tabular}{|c|c|c|c|c|c|c|}
\hline 
Compound & Ox. $\mbox{LaTiO}_3$$^{a}$ & Sr. $\mbox{LaTiO}_3$$^{b}$ & $\mbox{LaTiO}_3$ (10 u.c.)/$\mbox{SrTiO}_3$  & $\mbox{LaTiO}_3$ (15 u.c.)/$\mbox{SrTiO}_3$  & La. STO $^{c}$ & La. STO $^{c}$
\tabularnewline
\hline
\hline 
$A (\Omega cm/K^{2})$ & $8\times10{}^{-9}$ & $2\times10{}^{-9}$ & $3.6\times10{}^{-7}$  &  $1.3\times10{}^{-7}$  & $5\times10{}^{-7}$ & $3\times10{}^{-8}$\tabularnewline
\hline
\end{tabular}
\caption{Comparison of $A$ parameters (given in $\Omega cm/K^{2}$) measured in the 10 u.c. and 15 u.c. $\mbox{LaTiO}_3$/$\mbox{SrTiO}_3$ samples (taking the thickness $d^{10u.c.}$=12 nm, $d^{15u.c.}$=13.5 nm), with data obtained from literature on thin films and crystals for similar doping : $^{a}$ oxygen doped $\mbox{LaTiO}_3$
by Taguchi et al\cite{taguchi} ; $^{b}$ Sr doped $\mbox{LaTiO}_3$
by Tokura et al\cite{tokura} ; $^{c}$ Two different La doped $\mbox{SrTiO}_3$
by Okuda et al\cite{okuda}. }
\end{table}

In summary, we have measured the electronic transport properties of $\mbox{LaTiO}_3$/$\mbox{SrTiO}_3$ hetero-structures. The samples display a metallic behaviour and  a superconducting transition is observed at low temperature. Our analysis shows that  a 2-DEG is formed at the interface which extends mostly in the $\mbox{SrTiO}_3$  substrate, in agreement with the electronic reconstruction scenario\cite{okamoto04n}. This discovery opens the possibility to study the interplay between superconductivity and different electronic orders predicted to take place with ultra-thin $\mbox{LaTiO}_3$ films on $\mbox{SrTiO}_3$. According to our results in terms of carrier density, mobility and gas thickness, it should be possible to modulate significantly the behaviour of the 2-DEG by adjusting the number of charge carriers with an electrostatic gate.\\

The authors acknowledge L. Benfatto, M. Grilli, S. Caprara, C. Castellani, C. Di Castro for useful discussions and L. Dumoulin for technical support. \\

\newpage

\textbf{\large Supplementary information}\\
\large

\textbf{Growth and characterization}

We have grown epitaxial layers of $\mbox{LaTiO}_3$ using
excimer laser based PLD on commercially
available (Crystak gmbh Germany) single crystal substrates of $\mbox{SrTiO}_3$ 
cut along (100) and (110) crystallographic directions. While the (100)
subtrates were given a buffered HF treatment to expose $\mbox{TiO}_2$ terminated
surface, the (110) plane has Sr, Ti and oxygen ions on one plane and
hence HF treatment is irrelevant in this case. The substrates were
glued to the heater block of the PLD system and heated in oxygen pressure
of 200 mTorr in the temperature range of 850 to 950 $^\circ$C for one hour
to realize surface reconstruction.  This process has been used
routinely to grow epitaxial films and heterostructures of $YBa_2Cu_3O_{6+x}$ and
hole doped manganites. The source of $\mbox{LaTiO}_3$  is a stoichiometric
sintered target of 22 mm in diameter which was ablated in oxygen partial
pressure of $1\times10^{-4}$ Torr with energy fluence of $\sim1J/cm^{2}$ per pulse
at a repetition rate of 3 Hz to realize a gowth rate of 0.12 \AA/s. 
Under these conditions, the $\mbox{LaTiO}_3$  phase is grown on $\mbox{SrTiO}_3$ substrates, as shown by X-Rays diffraction patterns. On (100)$\mbox{SrTiO}_3$, a pure (100)$\mbox{LaTiO}_3$  film  is grown (supplementary figure 1). On (110)$\mbox{SrTiO}_3$ , a pure (110)$\mbox{LaTiO}_3$  is observed (supplementary figure 2). In both cases, film peaks are almost aligned with the corresponding substrate ones, which confirms the good orientations.\\

\normalsize
   \begin{figure}[h]
\centering
\includegraphics[width=12cm]{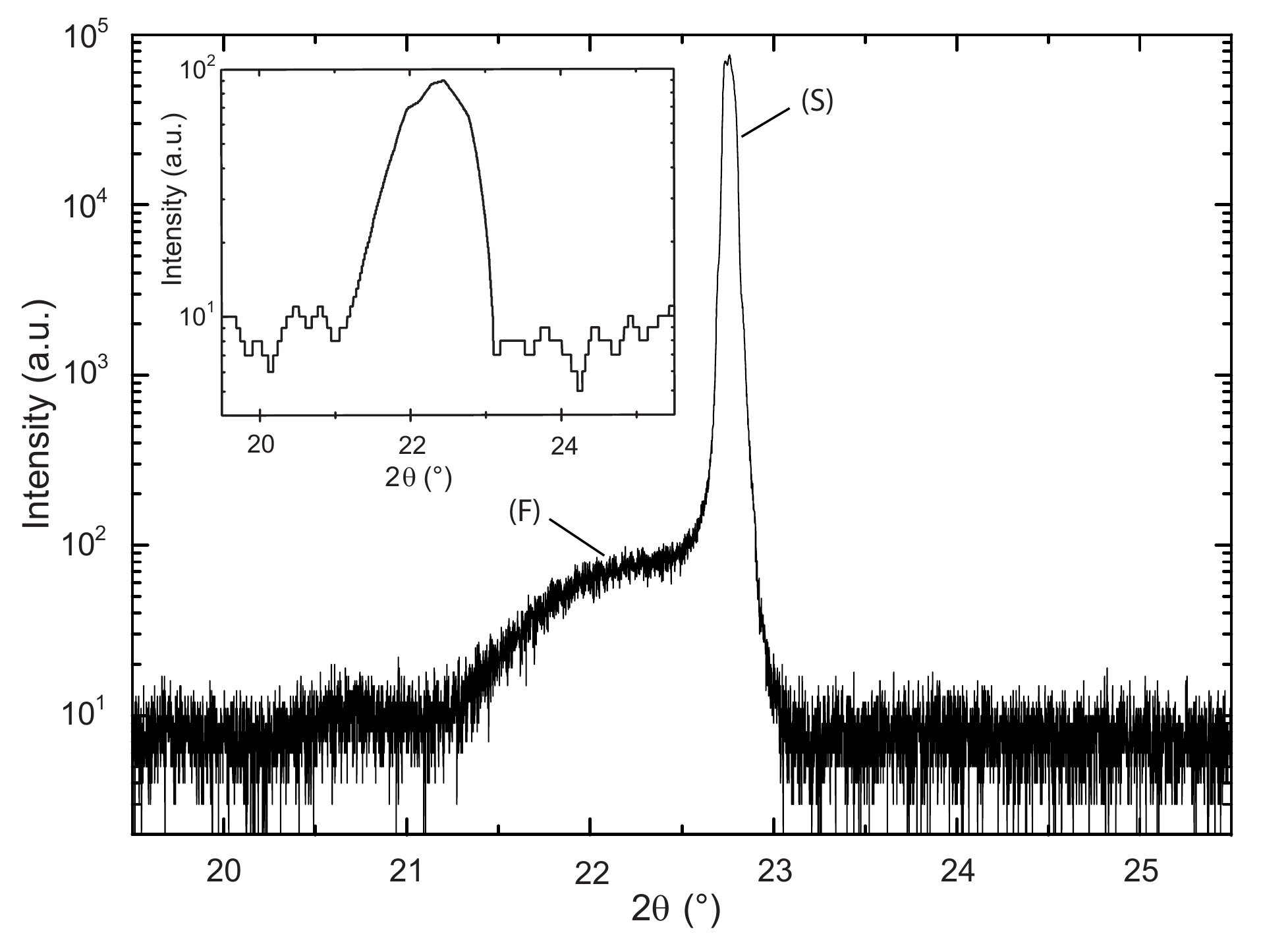}
\end{figure}
\newpage
Supplementary Figure 1. X-rays diffraction pattern ($\theta-2\theta$ scan) for $\mbox{LaTiO}_3$  films (F) deposited on (100)$\mbox{SrTiO}_3$ substrate. After subtracting the contribution of the substrate (S), the lattice parameter of the film is found to be 3.956 \AA  (inset), in good agreement with previous studies [Havelia et al. Journal of Crystal Growth \textbf{310}, 1985 (2008)] and close to the one reported in bulk $\mbox{LaTiO}_3$ (3.928 \AA) [Kestigian et al.,  J. Am. Chem. Soc. \textbf{76}, 6027 (1954)].\\

     \begin{figure}[h]
\centering
\includegraphics[width=12cm]{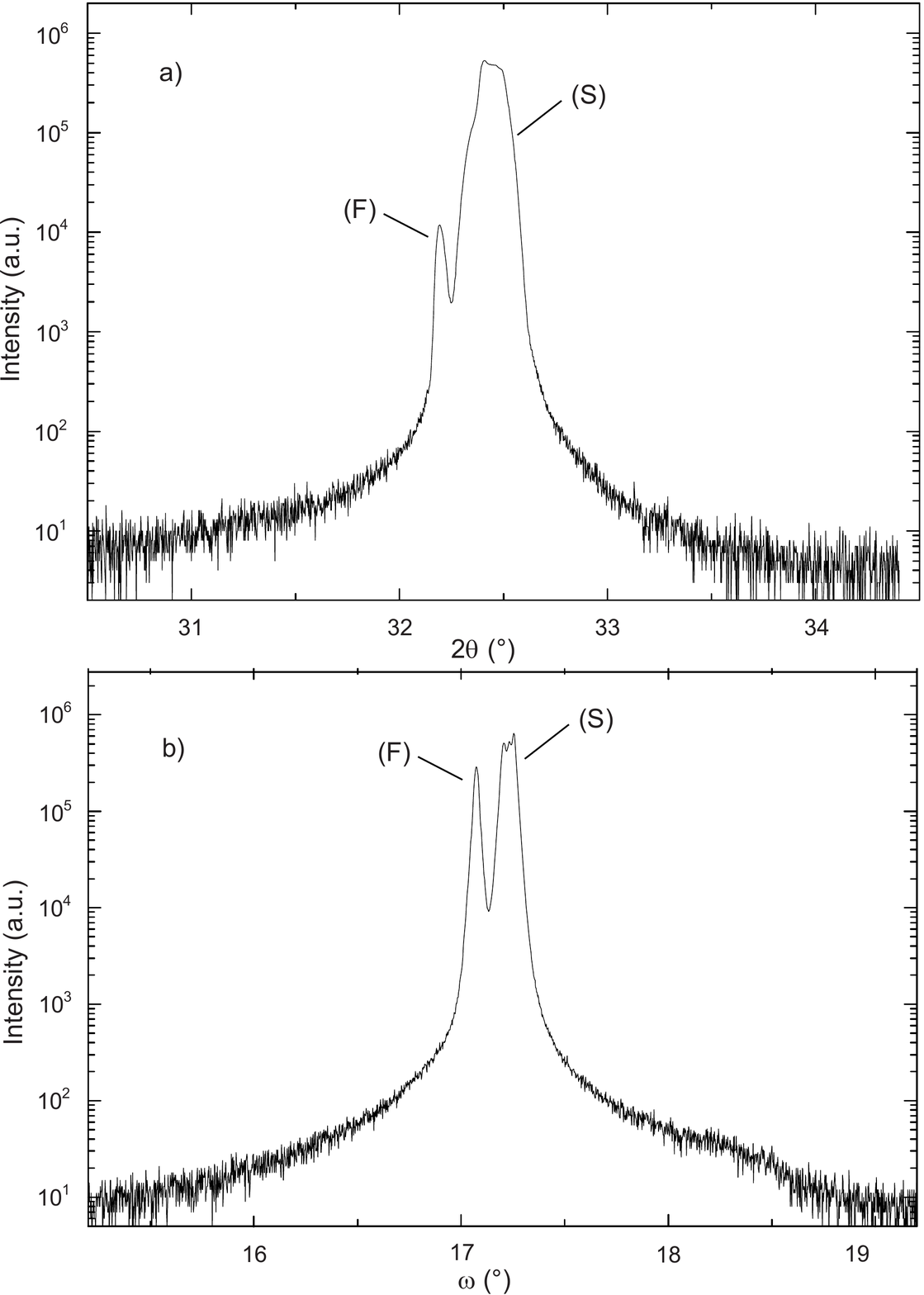}
\end{figure}

\newpage
Supplementary Figure 2.  a) X-rays diffraction pattern ($\theta$-2$\theta$ scan) around 32$^\circ$  of  $\mbox{LaTiO}_3$ films deposited on $\mbox{SrTiO}_3$(110). The (110) peak of the film (F) is observed at 2$\theta$=32.193 $^\circ$, close to the (110) peak of the substrate (S), which  corresponds to a $\mbox{LaTiO}_3$ lattice parameter of 3.928 \AA. b) X-rays diffraction rocking curve ($\omega$ scan) of the (110) peak of $\mbox{LaTiO}_3$ films on $\mbox{SrTiO}_3$(110). The typical width of the curve is around 0.1$^\circ$ showing a very good out-of-plane plane orientation of the layers.\\
 
     \begin{figure}[h]
\centering
\includegraphics[width=12cm]{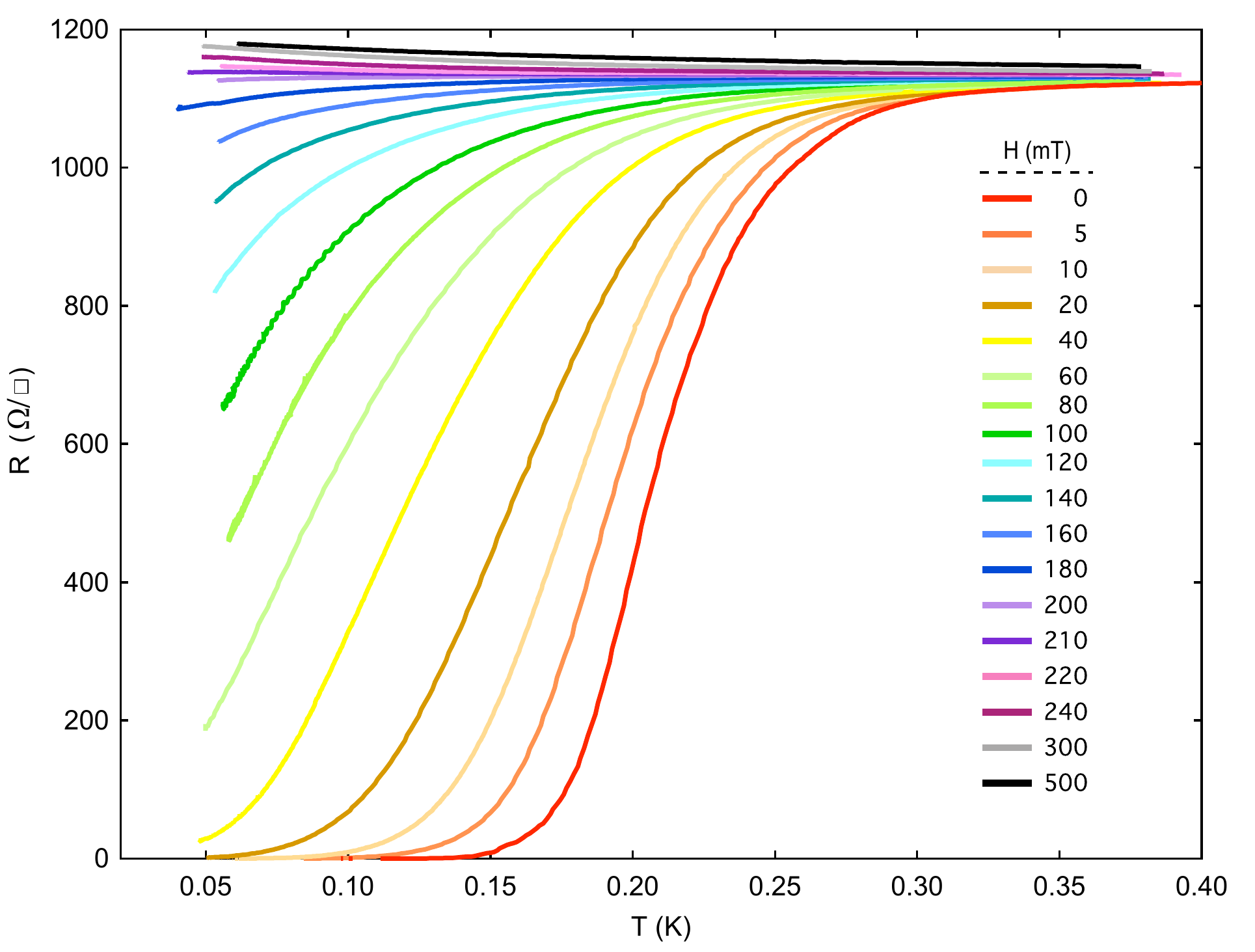}
\end{figure}
\newpage
Supplementary Figure 3. Sheet resistance of the 15 u.c. sample as a function of temperature for different values of the perpendicular magnetic field. 
\end{document}